\documentclass[reprint, NumberedRefs]{JASA-EL}

\makeatletter
\providecommand{\interlinepenalty@ltx}{\interlinepenalty}
\providecommand{\clubpenalty@ltx}{\clubpenalty}
\providecommand{\widowpenalty@ltx}{\widowpenalty}
\providecommand{\brokenpenalty@ltx}{\brokenpenalty}

\def\frontmatter@abstract@produce{%
  \par
  \begingroup
    \unvbox\absbox
  \endgroup
  \@ifx{\@empty\mini@notes}{}{\mini@notes\par}%
  \vskip6pt
  \normalsize
}
\makeatother

\usepackage{ragged2e}
\usepackage{xcolor}

\begin{document}

\title[JASA-EL/Sample JASA-EL Article]{Closed-form weak-wave propagation pressure amplitudes for arbitrary power-law attenuation}

\author{Esteban Avilés}
\email{esteban.aviles@pucp.edu.pe}
\affiliation{Laboratorio de Imágenes Médicas, Departamento de Ingeniería,  Pontificia Universidad Católica del Perú, Lima 15088, Peru}
\author{Michael Oelze}
\email{oelze@illinois.edu}
\affiliation{Beckman Institute for Advanced Science and Technology, Department of Electrical and Computer Engineering,  University of Illinois Urbana-Champaign, Urbana, Illinois 61801, USA}
\author{Roberto Lavarello}
\email{lavarello.rj@pucp.edu.pe}
\author{Andres Coila}
\email{acoila@pucp.edu.pe}
\affiliation{Laboratorio de Imágenes Médicas, Departamento de Ingeniería,  Pontificia Universidad Católica del Perú, Lima 15088, Peru}

\date{25 August 2026}

\preprint{Avilés et al., JASA-EL}

\begin{abstract}

Biological tissues exhibit power-law frequency dependence of the attenuation coefficient. We derive closed-form weak-wave expressions for fundamental, second-, and third-harmonic pressure amplitudes of monochromatic acoustic plane waves in homogeneous media with arbitrary power-law attenuation to extend classical expressions for quadratic attenuation. For linear frequency dependence of attenuation, one-dimensional simulations using the k-Wave k-space pseudospectral solver yielded 35--42\% (classical) and 0--2\% (proposed) normalized root-mean-square errors for fundamental-band depletion. This extension supports improved fundamental-band \textit{B/A} estimation for quantitative ultrasound tissue characterization. A proof-of-concept through-transmission application in phantoms with nearly linear attenuation yielded experimental \textit{B/A} estimation errors of 23.7--52.7\% (classical) and 4.0--5.9\% (proposed).

\end{abstract}

\maketitle

\section{Introduction}

\label{sec:Introduction}

Finite-amplitude wave propagation of a monochromatic plane wave redistributes energy from the fundamental frequency to harmonics due to the nonlinearity of the medium. For weak waves in homogeneous lossy fluids, characterized by a Gol'dberg number, $\Gamma$, less than 1, an analytical description derived from the Burgers' equation yields closed-form expressions for fundamental and higher harmonic amplitudes under classical thermoviscous dissipation, for which the attenuation coefficient (AC) is proportional to the square of angular frequency, $\alpha(\omega)\propto\omega^2$ \cite{hamilton1998}. However, in biological tissues, the AC is better described by the power-law attenuation model $\alpha(\omega)=\alpha_0|\omega|^y$, where $\alpha_0$ is expressed in Np/m/(rad/s)$^y$ (or its equivalent dB/cm/MHz$^y$), and $y$ is the attenuation power-law exponent. Hereafter, linear and quadratic attenuation denote the cases $y=1$ and $y=2$, respectively. For many soft biological tissues, $y$ commonly lies between 1 and 2 over limited frequency ranges and is often close to 1, rather than following quadratic frequency dependence \cite{cobbold2006}. Power-law attenuation in biological tissues can arise from a distribution of viscothermal and viscoelastic relaxation mechanisms, while scattering from tissue microstructure provides an additional frequency-dependent contribution \cite{jongen1986}.

This difference is relevant for finite-amplitude nonlinearity parameter, \textit{B/A}, estimation. \textit{B/A} has been investigated as a quantitative tissue-characterization parameter because it is sensitive to changes in tissue fat content \cite{panfilova2021}. A closed-form lossy-medium expression for second-harmonic generation exists without assuming a specific frequency dependence of the AC \cite{thuras1935}. Many established transmission, insert-substitution, and echo-mode strategies build on analytical forward models of this type for second-harmonic-based \textit{B/A} estimation \cite{gong1989,dong1999,varray2011}. Fundamental-band approaches have also been explored in multi-energy frameworks \cite{fatemi1996}, but these methods require strong planar reflectors or wire targets \cite{panfilova2021}. More recent formulations address these practical limitations \cite{coila2025}; however, they still use expressions derived for quadratic thermoviscous loss. To the best of our knowledge, corresponding weak-wave closed-form expressions for the fundamental band or higher harmonics beyond the second harmonic in media with arbitrary power-law attenuation have not been reported. Because these approaches estimate \textit{B/A} from measured spectral amplitudes, their implementation requires an explicit forward relation between the measured amplitude and \textit{B/A}. Closed-form expressions are therefore useful for direct \textit{B/A} estimation strategies.

The power-law frequency dependence of the AC in soft tissues has motivated formulations beyond the quadratic attenuation case \cite{szabo1994}. In particular, fractional models have been developed to describe arbitrary power-law attenuation in lossy media including nonlinear extensions of Burgers-type equations \cite{chen2004}. However, they do not develop compact weak-wave closed-form expressions for the spectral amplitudes of a monochromatic plane wave. Their use for \textit{B/A} estimation would therefore require solving the propagation problem rather than evaluating an explicit spectral-amplitude relation.

In this article, closed-form weak-wave expressions are derived for the fundamental frequency and both second and third harmonics of a monochromatic plane wave in a homogeneous medium with arbitrary power-law attenuation, which extends the weak-wave solution for quadratic thermoviscous loss. The main contribution is therefore a compact analytical forward model that retains the weak-wave structure of the classical Burgers solution while incorporating the harmonic-dependent attenuation implied by arbitrary power-law attenuation. The present derivation assumes this weak-wave regime, consistent with the quasi-linear conditions commonly stated in finite-amplitude characterization of tissue-like media and phantoms \cite{hamilton1998}. The accuracy of the pressure field predictions from the analytical expressions derived in this work was first assessed by one-dimensional k-Wave simulations of homogeneous plane-wave propagation for AC values at the fundamental frequency from 2 to 6 dB/cm and power-law exponents $1\leq y \leq 2$. The fundamental-band expression was then evaluated in through-transmission phantom \textit{B/A} estimation in both simulations and experiments as a proof-of-concept application to determine whether it reduces bias relative to the classical model with quadratic attenuation.

\section{Theory}

A one-dimensional, progressive, finite-amplitude plane wave is considered in a homogeneous medium with arbitrary power-law attenuation and expressed in retarded time, $\tau=t-x/c_0$, where $t$ is time (s), $x$ is propagation distance (m), and $c_0$ is the small-signal speed of sound (m/s). For the derivation, $\omega$ is expressed in rad/s and $\alpha(\omega)$ in Np/m. For the source angular frequency $\omega_0$, let $\lambda_n=n^y$, such that $\alpha(n\omega_0)=\lambda_n\alpha(\omega_0)$. Under monochromatic excitation at angular frequency $\omega_0$ and within the weak-wave regime, the corresponding pressure field, $p(x,\tau)$ (Pa), can be modeled with a Burgers-type equation such that the classical thermoviscous loss term is replaced by a fractional Laplacian loss operator associated with $\alpha(\omega)$ \cite{chen2004}.

\begin{equation}
\frac{\partial p}{\partial x}
+\alpha_0
\left(-\frac{\partial^2}{\partial \tau^2}\right)^{y/2} p
=
\frac{\beta}{\rho_0 c_0^3}\,p\,\frac{\partial p}{\partial \tau},
\label{eq:model_dim}
\end{equation}

\noindent with source condition $p(0,\tau)=p_0\sin(\omega_0\tau)$. Here, $p_0$ is the peak source pressure amplitude (Pa), $\rho_0$ is the equilibrium density (kg/m\textsuperscript{3}), and $\beta=1+B/(2A)$ is the coefficient of nonlinearity. The fractional operator in Eq.~\eqref{eq:model_dim} is defined in the $\omega$ domain after applying the Fourier transform with respect to $\tau$, such that $\mathcal{F}_{\tau}\{(-\partial^2/\partial \tau^2)^{y/2}p\}=[-(i\omega)^2]^{y/2}\widehat{p}=|\omega|^y\widehat{p}$, where $\widehat{p}(x,\omega)$ is the Fourier transform of $p(x,\tau)$ with respect to $\tau$. Thus, the operator has angular-frequency Fourier symbol $|\omega|^y$. When $y=2$, Eq.~\eqref{eq:model_dim} reduces to the classical Burgers equation for quadratic thermoviscous loss \cite{hamilton1998}. It is convenient to introduce the lossless shock formation distance, $\bar{x}=\rho_0 c_0^3/[\beta\omega_0 p_0]$, where $\bar{x}$ is in meters and its relation to $\Gamma$ is given by $\Gamma=[\alpha(\omega_0)\bar{x}]^{-1}=\beta\omega_0 p_0/[\rho_0 c_0^3\alpha(\omega_0)]$. Then, the propagation distance, retarded time, and pressure are made dimensionless by defining $\eta=\alpha(\omega_0)x$, $\theta=\omega_0\tau$, and $P=p/p_0$, respectively. With these definitions, Eq.~\eqref{eq:model_dim} becomes

\begin{equation}
\frac{\partial P}{\partial \eta}+D_y P
=
\Gamma P\frac{\partial P}{\partial \theta},
\qquad
P(0,\theta)=\sin\theta,
\qquad
D_y\equiv\left(-\frac{\partial^2}{\partial \theta^2}\right)^{y/2},
\label{eq:model_nd}
\end{equation}

\noindent where $D_y$ is the fractional operator. Under nonlinear propagation, a monochromatic source generates harmonic components at integer multiples of $\omega_0$. For these harmonic components, the operator satisfies $D_y\sin(n\theta)=\lambda_n\sin(n\theta)$. Therefore, the AC of the $n$-th harmonic scales as $\lambda_n$. For $\Gamma<1$, from a regular perturbation expansion of Eq.~\eqref{eq:model_nd} in powers of $\Gamma$, i.e., $P=P^{(0)}+\Gamma P^{(1)}+\Gamma^2P^{(2)}+\mathcal{O}(\Gamma^3)$, the zeroth-order solution can be found to be $P^{(0)}=e^{-\eta}\sin\theta$. At order $\mathcal{O}(\Gamma)$, the nonlinear term is $P^{(0)}\partial P^{(0)}/\partial\theta$, which generates the second harmonic. At order $\mathcal{O}(\Gamma^2)$, the nonlinear terms $P^{(0)}\partial P^{(1)}/\partial\theta$ and $P^{(1)}\partial P^{(0)}/\partial\theta$ generate a correction to the fundamental band and the first contribution at the third harmonic. Retaining terms through this order is sufficient to describe the fundamental band and the first two generated harmonics \cite{keck1960}, i.e.,

\begin{equation}
P(\eta,\theta)
=
a_1(\eta)\sin\theta
+a_2(\eta)\sin(2\theta)
+a_3(\eta)\sin(3\theta)
+\mathcal{O}(\Gamma^3),
\label{eq:harmonic_ansatz}
\end{equation}

\noindent where $a_n(\eta)$ is the dimensionless amplitude coefficient of the $n$-th harmonic. Hence, through $\mathcal{O}(\Gamma^2)$, $a_1=\mathcal{O}(1)$ with an $\mathcal{O}(\Gamma^2)$ correction, $a_2=\mathcal{O}(\Gamma)$, and $a_3=\mathcal{O}(\Gamma^2)$. The source conditions are $a_1(0)=1$ and $a_2(0)=a_3(0)=0$. With this ordering, the only terms in $\Gamma P\partial P/\partial\theta$ that contribute through $\mathcal{O}(\Gamma^2)$ are proportional to $\Gamma a_1^2$ or $\Gamma a_1a_2$. Substituting Eq.~\eqref{eq:harmonic_ansatz} into Eq.~\eqref{eq:model_nd}, arranging terms

with the same sine harmonic, and retaining terms through $\mathcal{O}(\Gamma^2)$ gives

\begin{equation}
\begin{aligned}
\frac{d a_1}{d \eta} + a_1 = -\frac{\Gamma}{2}a_1a_2, \qquad
\frac{d a_2}{d \eta} + \lambda_2 a_2 = \frac{\Gamma}{2}a_1^2, \qquad
\frac{d a_3}{d \eta} + \lambda_3 a_3 = \frac{3\Gamma}{2}a_1a_2.
\end{aligned}
\label{eq:a_odes}
\end{equation}

Consistent with this ordering, Eq.~\eqref{eq:a_odes} is solved by introducing $a_1(\eta)=e^{-\eta}+\Gamma^2b_1(\eta)$, $a_2(\eta)=\Gamma b_2(\eta)$, and $a_3(\eta)=\Gamma^2 b_3(\eta)$, where $b_1$, $b_2$, and $b_3$ are auxiliary functions. Substitution into Eq.~\eqref{eq:a_odes} gives

\begin{equation}
\begin{aligned} \frac{d b_1}{d\eta}+b_1 &= -\frac{1}{2}e^{-\eta}b_2, \qquad \frac{d b_2}{d\eta}+\lambda_2 b_2 = \frac{1}{2}e^{-2\eta}, \qquad \frac{d b_3}{d\eta}+\lambda_3 b_3 = \frac{3}{2}e^{-\eta}b_2,
\end{aligned}
\end{equation}

\noindent with $b_1(0)=b_2(0)=b_3(0)=0$. Solving these equations gives

\begin{align}
b_1(\eta)
&=
-\frac{e^{-\eta}}{4(\lambda_2-2)}
\left[
\frac{1-e^{-2\eta}}{2}
-\frac{1-e^{-\lambda_2\eta}}{\lambda_2}
\right],
\label{eq:b1}
\\
b_2(\eta)
&=
\frac{e^{-2\eta}-e^{-\lambda_2\eta}}{2(\lambda_2-2)},
\label{eq:b2}
\\
b_3(\eta)
&=
\frac{3}{4(\lambda_2-2)}
\left[
\frac{e^{-3\eta}-e^{-\lambda_3\eta}}{\lambda_3-3}
-
\frac{e^{-(1+\lambda_2)\eta}-e^{-\lambda_3\eta}}{\lambda_3-1-\lambda_2}
\right].
\label{eq:b3}
\end{align}

The expressions in Eqs.~\eqref{eq:b1}--\eqref{eq:b3} are referred to as the Extended Burgers expressions (EBE). At $y=1$, the limiting EBE pressure amplitudes are obtained from the limit $y\to1$ and can be written as

\begin{equation}
p_1(x) = p_0 e^{-\eta} \left [ 1- \frac{\Gamma^2}{16}  \left ( 1-\left(1+2\eta \right)e^{-2\eta} \right ) \right ],
\quad
p_2(x) = p_0\frac{\Gamma}{2} \eta e^{-2\eta},
\quad
p_3(x)=\frac{3}{8}p_0 \Gamma^2 \eta^2 e^{-3\eta}.
\label{eq:y1_ebe}
\end{equation}

In the quadratic attenuation case, $y=2$, Eqs.~\eqref{eq:b1}--\eqref{eq:b3} recover the classical weak-wave Burgers expressions of Keck and Beyer~\cite{keck1960}. These expressions are referred to here as the classical Burgers expressions (CBE) and are

\begin{equation}
\begin{aligned}
p_1(x) &= p_0 e^{-\eta} \left [ 1- \frac{\Gamma^2}{32}  \left ( 1-e^{-2\eta} \right )^2 \right ],
\quad
p_2(x) &= p_0\frac{\Gamma}{4}e^{-2\eta} \left( 1-e^{-2\eta} \right),
\quad \\
p_3(x)&=p_0 \frac{\Gamma^2}{32}e^{-3\eta} \left( 1-e^{-2\eta} \right)^2 \left( 2+e^{-2\eta} \right).
\end{aligned}
\label{eq:y2_cbe}
\end{equation}

In Eqs.~\eqref{eq:y1_ebe} and \eqref{eq:y2_cbe}, $p_n$ denotes the peak pressure amplitude of the $n$-th harmonic at distance $x$. Eq.~\eqref{eq:b2} is also consistent with the finite-amplitude result reported by Thuras \textit{et al.} \cite{thuras1935}, upon identifying $\eta=\alpha(\omega_0)x$ and using $\alpha(2\omega_0)=\lambda_2\alpha(\omega_0)$. The expressions do not impose a specific frequency range, the formulation is applicable to acoustic frequencies in general under the assumptions stated above.

\section{Methodology}

\subsection{Plane-wave numerical validation}

The closed-form expressions were first validated against one-dimensional k-Wave simulations of homogeneous progressive plane-wave propagation \cite{treeby2012}. The k-Wave solver computes pressure fields by solving coupled first-order acoustic equations using a k-space pseudospectral method. In this study, k-Wave was used to generate the temporal pressure field along the propagation axis, which was then processed to extract the fundamental, second-harmonic, and third-harmonic amplitudes. Only power-law absorption was included in the simulations, whereas causal dispersion was omitted to match the assumptions of the closed-form expressions. The source was a sinusoidal wave with frequency $f_0=5$ MHz ($\omega_0=2\pi f_0$) that covered the entire simulated time domain. The medium parameters were fixed at $c_0=1500$ m/s, $\rho_0=1000$ kg/m\textsuperscript{3}, and \textit{B/A} = 6. The spatial grid step was $\Delta x=10$ $\mu$m, corresponding to 30 points per wavelength at $f_0$. The maximum grid-supported frequency, given by $f_{\max}=c_0/(2\Delta x)$, was 75 MHz, well above the third harmonic analyzed in this work.

Validation was performed over a sweep of two parameters: $\alpha(\omega_0)$ values from 2 to 6 dB/cm in steps of 0.5 dB/cm, and $y$ values from 1 to 2 in steps of 0.1. The AC range was chosen so that $\Gamma$ remained below 1 in all reported cases, consistent with the weak-wave regime assumed in the derivation. For each parameter pair, the simulated pressure field generated by k-Wave was taken as ground truth (GT) and compared with two analytical models: the CBE model, evaluated with quadratic attenuation, and the EBE model, evaluated with an arbitrary power-law exponent. For both scenarios, the AC was matched at $\omega_0$. The quantities compared were the fundamental-band depletion term, $\Delta p_1=\nu p_{1,L}-p_{1,H}$, and the second- and third-harmonic amplitudes $p_2$ and $p_3$. Here, $\nu$ is a source pressure scaling factor, and the subscripts $L$ and $H$ denote low- and high-pressure source levels, respectively. Under linear scaling, $\Delta p_1$ cancels the linear fundamental contribution and isolates more clearly the changes in $p_1$ due to nonlinearity. For the fundamental band, $p_{0,L}$ and $p_{0,H}$ were set to 80 and 400 kPa, respectively. For $p_2$ and $p_3$, the high-pressure case was used.

The amplitudes $p_1$, $p_2$, and $p_3$ were extracted from the simulated time signals by zero-phase FIR filtering over 0.2 MHz bands centered at $f_0$, $2f_0$, and $3f_0$, respectively, followed by Hilbert-envelope detection. The harmonic envelopes were averaged over the steady-state interval in a window of approximately 20 $\mu$s, which corresponds to 100 cycles at $f_0$. This window had a 1.2\% fractional bandwidth at --6 dB. Each FIR filter had order 6000, with a bandwidth of 0.2 MHz, sufficient for this application. For each quantity $q\in\{\Delta p_1,p_2,p_3\}$, agreement between each analytical model, $M\in\{\mathrm{CBE},\mathrm{EBE}\}$, and GT was quantified using the normalized root-mean-square error (NRMSE), expressed as a percentage:

\begin{equation}
\varepsilon_{M,q}(y,\alpha_j)[\%] = 100 \frac{\left\|q_{M}(\mathcal{Z})-q_{\mathrm{GT}}(\mathcal{Z})\right\|_2}{\left\|q_{\mathrm{GT}}(\mathcal{Z})\right\|_2},
\label{eq:relative_l2_error}
\end{equation}

\noindent where $\mathcal{Z}$ denotes the 0.5--6 cm depth interval, $q_M$ is the analytical profile from model $M$, $q_{\mathrm{GT}}$ is the k-Wave profile, and $\alpha_j$ is the $j$-th value of $\alpha(\omega_0)$ in the AC sweep. The first 0.5 cm was excluded because $\Delta p_1$, $p_2$, and $p_3$ are small near the source, making the NRMSE more sensitive to numerical noise than to model mismatch. For each value of $y$, the mean, minimum, and maximum NRMSE values for CBE and EBE were computed across the AC sweep. For each pair $(y,\alpha_j)$, the error reduction was defined in percentage points (pp) as the CBE error minus the EBE error. Axial-profile error reduction was also computed as the difference between the CBE and EBE pointwise absolute percentage errors at each depth.

\subsection{Through-transmission \textit{B/A} estimation in samples}

A second study evaluated the $p_1$ expression for through-transmission \textit{B/A} estimation in tissue-mimicking phantoms (TMPs) and corn oil. The corn oil sample was added to verify agreement between the CBE and EBE schemes under $y\approx 2$ case. The study combined 3D simulations and sample measurements using a common processing pipeline. In the experiments with the TMPs, transmission was performed with a 17-element sub-aperture of an Ultrasonix (Analogic, Salem, NH) L9-4/38 linear array, driven in phase, with pitch 0.3 mm and element height 5 mm. The transmitted waveform was centered at 5 MHz with an 11.6\% fractional bandwidth at $-6$ dB. Reception was performed with an axially aligned 5 MHz single-element transducer with a 6.35 mm aperture (ndtXducer, Englewood, FL) placed behind the medium. A Daedal positioning system (Parker Hannifin Corp., Cleveland, OH) was used to position the receive transducer. Signal transmission was driven with a Verasonics Vantage 128 system (Verasonics Inc., Kirkland, WA), and the received waveform from the single-element transducer was recorded with an oscilloscope. For measurements of the corn oil sample, the same experimental configuration was used, except that transmission was performed with a Verasonics L11-4v linear array. In the 3D simulations, the transmit aperture was modeled as a 17-element flat aperture with 0.3 mm pitch, driven in phase, and the receiver was modeled as an axially aligned single-element 6.35 mm circular aperture.

Two of the TMPs were made from agarose, $n$-propanol, condensed milk, and water, with glass bead scatterers uniformly placed at random \cite{wear2005}. An oil-in-agar sample phantom was also fabricated, with a formulation of 2\% agar, 3\% gelatine, 25\% corn oil, and 1\% emulsificant. The AC was characterized by the insertion-loss approach \cite{kuc1979} using a Panametrics 5072PR pulser/receiver over 3--7 MHz for the TMPs and 2--8 MHz for corn oil. Sound speed was obtained by the same approach, and \textit{B/A} was measured separately using the method of Dong \textit{et al.} \cite{dong1999}. Hereafter, S1 and S2 denote the agarose phantoms, S3 denotes the oil-in-agar phantom, and S4 denotes the corn oil sample. Their properties were: for S1, \textit{B/A} = 6.78, $c=1538$ m/s, $y=1.00$, and $\alpha_0=0.70$ dB/cm/MHz$^{1.00}$; for S2, \textit{B/A} = 6.75, $c=1534$ m/s, $y=1.18$, and $\alpha_0=0.34$ dB/cm/MHz$^{1.18}$; for S3, \textit{B/A} = 9.21, $c=1470$ m/s, $y=1.00$, and $\alpha_0=0.83$ dB/cm/MHz$^{1.00}$; and for S4, \textit{B/A} = 10.60, $c=1455$ m/s, $y=1.96$, and $\alpha_0=0.05$ dB/cm/MHz$^{1.96}$. The corresponding homogeneous simulation media were assigned the same acoustic properties as the physical samples.

In the experimental implementation, S1, S2, and S3 had a thickness of 4 cm, whereas S4 had a thickness of 3 cm. For S1--S3, the low and high drive-voltage levels were 8 V and 40 V, respectively. The incident pressures were measured with a needle hydrophone near the transmit surface (closer than 1 mm), resulting in 89.4 and 486 kPa, respectively, giving $\nu \approx 5.44$. For S4, the low and high drive-voltage levels were 8 V and 20 V, corresponding to incident pressures of 93.8 and 230 kPa, respectively, and $\nu \approx 2.45$. Using the measured $p_{0,H}$ at 5 MHz, the $\Gamma$ values were 0.46 for S1, 0.71 for S2, 0.56 for S3, and 0.97 for S4. For both simulations and experiments, the received $p_{1,L}$ and $p_{1,H}$ values from the low- and high-pressure signals were obtained by filtering over $f_0 \pm 1$ MHz followed by envelope detection. The normalized depletion magnitude was defined as $d=\Delta p_1/p_{1,L}$. Then, using the simulated or measured pressure data as input, the nonlinearity coefficient of the sample was estimated from Eq.~\eqref{eq:b1} and in terms of $d$ as

\begin{equation}
\widehat{\beta}
=
\frac{\rho_0 c^3\alpha(\omega_0)}{\omega_0p_{0,L}}
\sqrt{
\frac{e^{-\eta} d}
{-b_1(\eta)\left[\nu^3-\nu+d\right]}
}.
\label{eq:tt_inversion}
\end{equation}

In Eq.~\eqref{eq:tt_inversion}, $b_1(\eta)$ is evaluated with either the CBE or EBE models. The corresponding $\widehat{\text{\textit{B/A}}}$ values were obtained from $\widehat{\text{\textit{B/A}}}=2(\widehat{\beta}-1)$. For the numerical implementation, 3D k-Wave simulations were performed in homogeneous media using power-law absorption only, without dispersion \cite{treeby2012}. For each sample, the sample thickness was varied from 0.5 to 6 cm in steps of 0.5 cm, with the measured pressures applied at the transmit aperture. Accuracy was assessed against the characterized \textit{B/A} using absolute relative percentage error.

\section{Results}

\subsection{Plane-wave numerical validation}

\begin{figure}[h!]

    \centering

    \includegraphics[width=\linewidth]{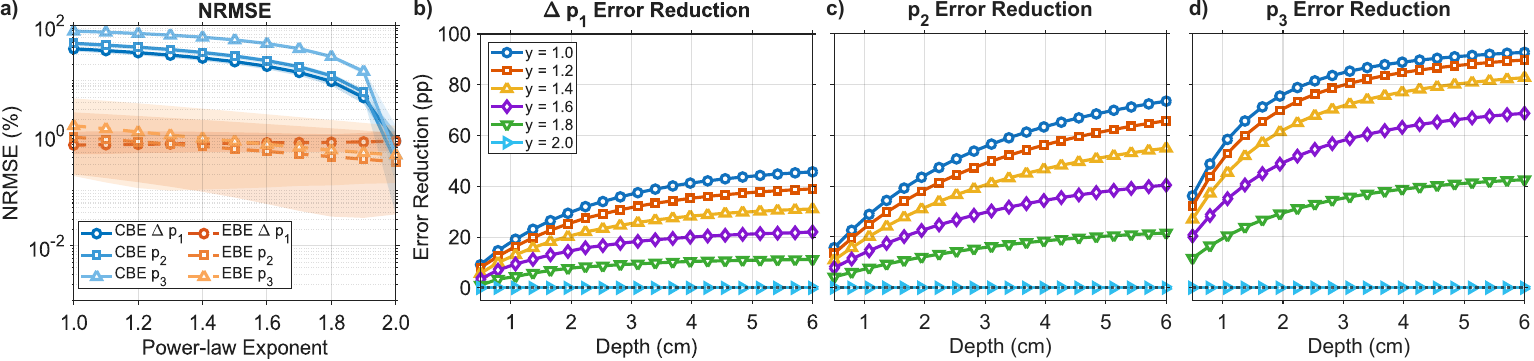}

    \caption{Plane-wave validation of CBE and the proposed EBE against one-dimensional homogeneous k-Wave simulations. (a) NRMSE for $\Delta p_1$, $p_2$, and $p_3$ as a function of $y$. Markers denote the mean error over the AC sweep, and shaded regions denote the corresponding minimum-to-maximum range. (b)--(d) Axial-profile error reduction when $\alpha(\omega_0)=3$ dB/cm for (b) $\Delta p_1$, (c) $p_2$, and (d) $p_3$ for the values of $y$ indicated in the legend. Error reduction is reported in pp as the CBE error minus the EBE error.}

    \label{fig:planewave_validation}

\end{figure}

Figure~\ref{fig:planewave_validation}(a) shows the NRMSE obtained with CBE and EBE in the one-dimensional homogeneous plane-wave simulations. The CBE NRMSE increased as $y$ decreased from the $y=2$ case for $\Delta p_1$, $p_2$, and $p_3$, across the tested exponent range. The shaded regions show the minimum-to-maximum range across the AC sweep. At $y=1$, the NRMSE values for $\Delta p_1$, $p_2$, and $p_3$ ranged from 35--42\%, 45--49\%, and 79--80\% using CBE, and 0--2\%, 0--3\%, and 0--5\% using EBE, respectively. The axial-profile error reduction curves for $\alpha(\omega_0)=3$ dB/cm showed the same trend in Figs.~\ref{fig:planewave_validation}(b)--(d). Error reduction increased with propagation distance and was largest for $y\to 1$ attenuation laws. The curves also showed an error reduction convergence that depended mainly on $y$. The largest error reductions were observed for $p_3$, followed by $p_2$ and $\Delta p_1$.

\subsection{Through-transmission \textit{B/A} estimation in samples}

\begin{figure}[h!]

    \centerline{\includegraphics[width=0.9\linewidth]{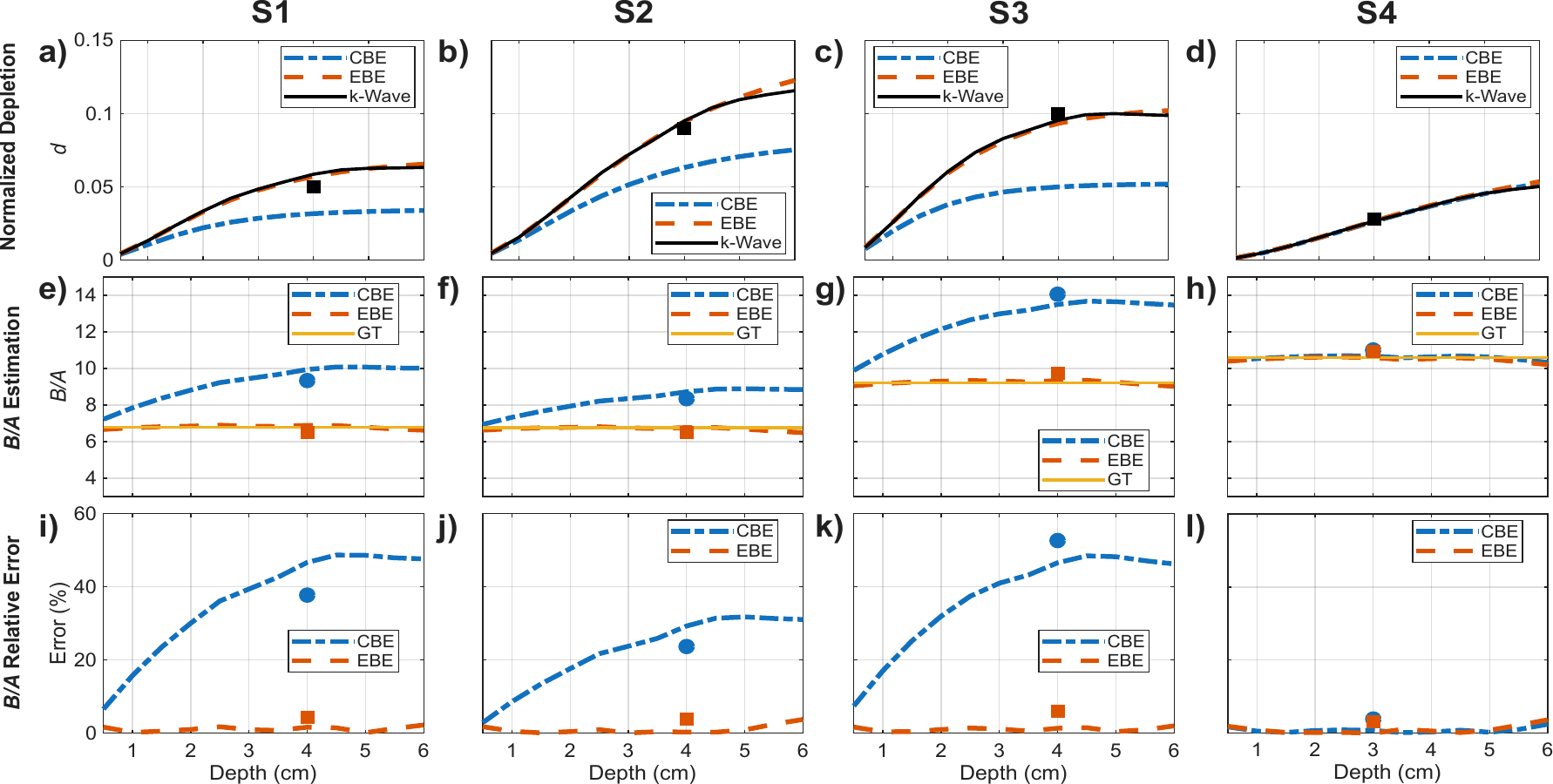}}

    \caption{Through-transmission proof-of-concept results for S1 (first column) to S4 (last column). (a)--(d) show CBE and EBE predictions of $d$ compared with k-Wave simulations. Black square markers denote the experimental $d$ values. (e)--(h) show the estimated \textit{B/A} as in Eq.~\eqref{eq:tt_inversion} while (i)--(l) show the relative errors. Blue circular and red square markers denote the corresponding CBE and EBE experimental estimates. The GT value for each sample is also shown.}

    \label{fig:tt_simulation}

\end{figure}

Figure~\ref{fig:tt_simulation} shows the through-transmission proof-of-concept for S1--S4 as the sample thickness was varied from 0.5 to 6 cm. Figs.~\ref{fig:tt_simulation}(a)--(d) compare $d$, obtained from simulated received pressures, with the CBE and EBE forward predictions. For S1--S3, the EBE prediction followed more closely the simulated $d$ curve than CBE, whereas for S4 the CBE and EBE predictions both agreed with the simulated curve. At 4 cm, simulation estimates using the CBE model were 9.95, 8.73, and 13.50 for S1, S2, and S3, corresponding to relative errors of 46.7\%, 29.3\%, and 46.6\%, respectively. Estimates with the EBE scheme were 6.89, 6.77, and 9.34, reducing the corresponding errors to 1.7\%, 0.3\%, and 1.4\%, respectively. For S4 at 3 cm, the CBE and EBE estimates were 10.68 and 10.59, corresponding to relative errors of 0.8\% and 0.1\%, respectively.

The experimental estimates for S1--S3 at 4 cm and S4 at 3 cm are also shown as single markers and follow the same overall trend. The CBE experimental estimates were 9.34, 8.35, and 14.06 for S1, S2, and S3, corresponding to relative errors of 37.8\%, 23.7\%, and 52.7\%, respectively. Estimation using the EBE scheme produced estimates of 6.49, 6.48, and 9.76, reducing the corresponding errors to 4.3\%, 4.0\%, and 5.9\%, respectively. Across S1--S3, the mean relative error was reduced from 38.1\% with CBE to 4.8\% with EBE. For S4, the CBE and EBE experimental estimates were 11.01 and 10.93, corresponding to relative errors of 3.9\% and 3.1\%, respectively.

\section{Discussion}

Closed-form weak-wave expressions for the fundamental, second, and third harmonics in homogeneous media with arbitrary power-law attenuation were derived. The main extensions are the $\Delta p_1$ term and $p_3$. The $p_2$ expression completes the low-order description and is consistent with Thuras \textit{et al.} \cite{thuras1935}. The analytical expressions and numerical validation included power-law absorption but omitted the associated causal dispersion \cite{waters2000}. For S1--S3, the Kramers--Kronig relation predicts increases in phase velocity of approximately 0.9--1.5 m/s. This frequency dependence could change the predicted fundamental depletion and consequently the \textit{B/A} estimates. The results also suggest that the approximation is less accurate as the weak-wave limit is approached. This is expected since the expressions retain terms only up to $\mathcal{O}(\Gamma^2)$, so higher-order terms should become more relevant as $\Gamma$ approaches unity. The expressions may become inaccurate in highly nonlinear media, such as liquids containing ultrasound contrast agents or other bubbly liquids. In such media, second-harmonic \textit{B/A} measurements may also contain systematic errors \cite{xia2019}.

The results from the one-dimensional simulations indicated that the assumption of a quadratic attenuation law in CBE was not sufficient when the AC followed a non-quadratic power law. Matching the AC at $\omega_0$ makes the CBE and EBE models equal only at the source frequency. However, the generated harmonics are attenuated according to the values of $\alpha(2\omega_0)$ and $\alpha(3\omega_0)$, which depend on $y$. Therefore, a model with quadratic attenuation cannot preserve the relative attenuation of the $p_1$, $p_2$, and $p_3$ when $y\neq2$. This explains why the differences between CBE and EBE increased as $y$ moved away from the $y=2$ case. Although the validation sweep covered $1\leq y\leq2$ to show the transition to the quadratic attenuation limit, the main practical interest is the lower-exponent range that is commonly encountered in biological tissues, with $y$ from 1.0 to 1.4 \cite{cobbold2006}, in which S1--S3 also fall. In this range, the axial-profile error reduction for $\alpha(\omega_0)=3$ dB/cm reached about 30--45 pp. These tended to approach a error reduction limit with propagation distance. This is expected if EBE is taken as the correct weak-wave model in the tested range, because the remaining CBE--EBE difference is mainly determined by the harmonic attenuation factors $\lambda_n$. Thus, the limiting value depends primarily on $y$, while $\alpha(\omega_0)$ mainly affects the distance over which that value is approached.

The through-transmission study evaluated whether the $p_1$ expression could be used in a more practical \textit{B/A} estimation setting. In matched simulations and physical measurements, EBE reduced bias relative to CBE for S1--S3, whereas both models produced similar estimates for S4, as expected when $y\approx 2$. Therefore, incorporating the power-law frequency dependence of the AC in the forward expression can reduce systematic error in fundamental-band \textit{B/A} estimation \cite{coila2025}. This is relevant for clinical translation because fundamental-band strategies can be more compatible with pulse-echo ultrasound than second-harmonic or custom probe approaches \cite{panfilova2021}.

Although the analytical model assumes monochromatic propagation, the finite-duration burst had an 11.6\% fractional bandwidth, within the \textless28\% narrowband range associated with lower bias in dual-energy fundamental-depletion \textit{B/A} estimation \cite{coila2025}. Since $p_{1,L}$ and $p_{1,H}$ share the same frequency band and aperture, their common pressure-independent linear diffraction factor cancels in $d$; pressure-dependent nonlinear beam evolution may still contribute, but the low EBE errors in the matched 3D simulations suggest a small net effect under the evaluated configurations. Remaining simulation--experiment differences could arise from unmodeled electromechanical transducer responses, particularly pressure-dependent components, or from the AC measurements, as they were not verified in the linear-propagation regime and nonlinear propagation can bias AC estimates \cite{dhooge1999}. The estimator also remains sensitive to $\nu$ and $p_0$ calibration; such errors could place $\Gamma$ closer to the weak-wave limit than indicated by the nominal values. These estimates could be improved in future work by adding reference-phantom compensation for common propagation and system factors that are not represented explicitly in the closed-form forward model \cite{coila2025}.

Although the proof-of-concept was evaluated in through-transmission, the expressions also apply to methods that estimate \textit{B/A} from the fundamental band where a weak-wave model estimates nonlinearity as a deviation from linear scaling. This approach has been used in finite-amplitude loss measurements \cite{fatemi1996} and adapted to pulse-echo estimation \cite{coila2025}. The present formulation extends the forward model to arbitrary power-law attenuation, reducing model mismatch when the attenuation power-law exponent is closer to 1. This correction may also be relevant to model-based harmonic-ratio approaches that include the fundamental band \cite{schrempf2026}. These applications remain prospective and require further validation.

\section{Conclusion}

Closed-form weak-wave expressions were derived for $p_1$, $p_2$, and $p_3$ of a monochromatic progressive plane wave in homogeneous media with arbitrary power-law attenuation. The formulation recovered the CBE expressions at $y=2$ and remained accurate for $y\approx 1$, for which the classical model resulted in strong bias in harmonic prediction. In through-transmission simulations and phantom measurements, the EBE scheme reduced \textit{B/A} bias in media with near-linear power-law attenuation. These results establish a compact forward model for weak-wave propagation in media with arbitrary power-law attenuation and support further validation of its use in depletion-based \textit{B/A} characterization and imaging.

\begin{acknowledgments}

This research was funded by the Consejo Nacional de Ciencia, Tecnología e Innovación Tecnológica (CONCYTEC) y el Programa Nacional de Investigación Científica y Estudios Avanzados (PROCIENCIA) under the contest E073-2025-01 ``Tesis de Pregrado y Postgrado en Ciencia, Tecnología e Innovación Tecnológica" (award number PE501099673-2025) and partially funded by the Escuela de Posgrado PUCP through the Marco Polo 2025-2 program.

\end{acknowledgments}

\section*{Author Declarations}

\noindent \textit{Conflict of Interest}

\noindent The authors have no conflicts to disclose.

\section*{Data Availability}

\noindent The data that support the findings of this study are available from the corresponding author upon reasonable request.

\bibliography{articlebib}

\end{document}